\documentclass[letterpaper, 10 pt, conference]{ieeeconf}  

\IEEEoverridecommandlockouts                              

 \usepackage{graphicx}
 \usepackage[vlined]{algorithm2e}
 \usepackage{amsmath}
 \usepackage{subfig}
 \usepackage{hhline}
 \usepackage{tabularx}

\title{\LARGE \bf
Weakly- and Semi-Supervised Probabilistic Segmentation and Quantification of Ultrasound Needle-Reverberation Artifacts to Allow Better AI Understanding of Tissue Beneath Needles
}

\author{Alex Ling Yu Hung$^{1}$, Edward Chen$^{2}$ and John Galeotti$^{2}$
\thanks{*This present work was sponsored in part by US Army Medical contracts W81XWH-19-C0083, W81XWH-19-C0101, and W81XWH-19-C-0020, and by a PITA grant from the state of Pennsylvania DCED C000072473.}
\thanks{$^{1}$Alex Ling Yu Hung is with Biomedical Engineering Department, Carnegie Mellon University, Pittsburgh PA 15213, USA
        {\tt\small lingyuh@andrew.cmu.edu}}%
\thanks{$^{2}$Edward Chen and John Galeotti are with Robotics Institute, Carnegie Mellon University, Pittsburgh PA 15213, USA {\tt\small edward2c@andrew.cmu.edu}, {\tt\small jgaleotti@cmu.edu}}%
}

\begin{document}

\maketitle
\thispagestyle{empty}
\pagestyle{empty}

\begin{abstract}

Ultrasound image quality has continually been improving. However, when needles or other metallic objects are operating inside the tissue, the resulting reverberation artifacts can severely corrupt the surrounding image quality. Such effects are challenging for existing computer vision algorithms for medical image analysis. Needle reverberation artifacts can be hard to identify at times and affect various pixel values to different degrees. The boundaries of such artifacts are ambiguous, leading to disagreement among human experts labeling the artifacts. We propose a weakly- and semi-supervised, probabilistic needle-and-reverberation-artifact segmentation algorithm to separate the desired tissue-based pixel values from the superimposed artifacts. Our method models the intensity decay of artifact intensities and is designed to minimize the human labeling error. We demonstrate the applicability of the approach and compare it against other segmentation algorithms. Our method is capable of differentiating between the reverberations from artifact-free patches as well as of modeling the intensity fall-off in the artifacts.  Our method matches state-of-the-art artifact segmentation performance and sets a new standard in estimating the per-pixel contributions of artifact vs underlying anatomy, especially in the immediately adjacent regions between reverberation lines. Our algorithm is also able to improve the performance downstream image analysis algorithms.

\end{abstract}

\section{INTRODUCTION}

\begin{figure}[thpb]
     \centering
     \subfloat{\includegraphics[width=0.11\textwidth]{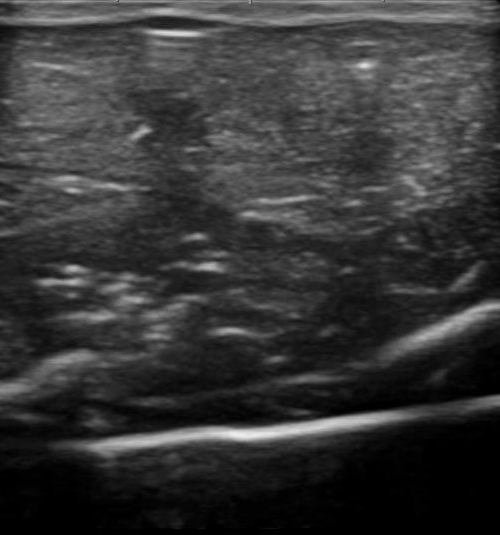}\label{ori}
     } 
     \subfloat{\includegraphics[width=0.11\textwidth]{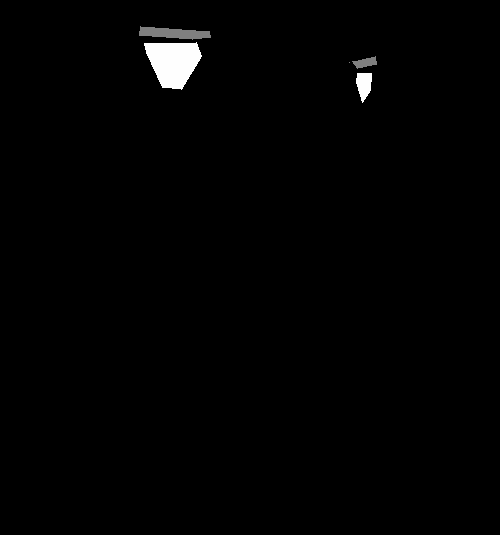}\label{lab1}
     } 
     \subfloat{\includegraphics[width=0.11\textwidth]{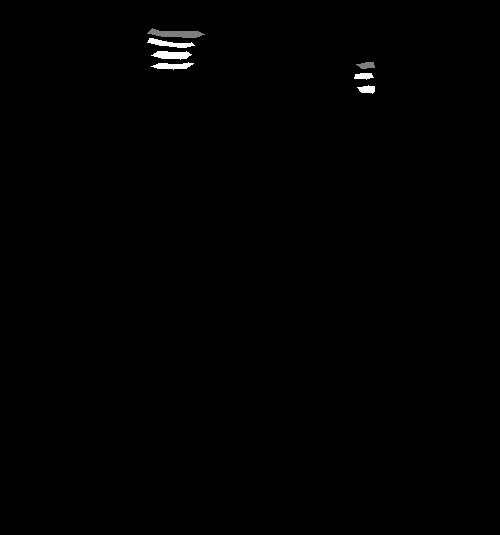}\label{lab2}
     } 
     \subfloat{\includegraphics[width=0.11\textwidth]{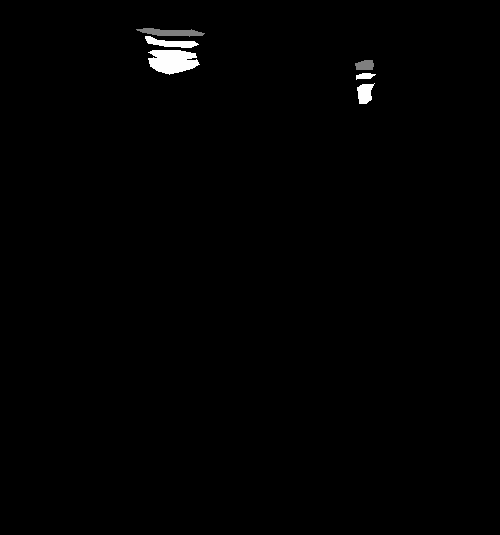}\label{lab3}
     }
     \caption{The first image from the left displays the original image, whereas the rest are examples of different labels from different annotators on the same chicken breast image. The gray labels are needles and the white labels are reverberation artifacts.  Annotators agree on the location of needles and reverberation artifacts, but disagree on the boundaries and the pixels between each reverberation.}
     \label{exa}
\end{figure}

Ultrasound imaging is widely used globally because it is a low-cost, safe, and fast imaging technique. Its real-time operation is perfect for monitoring needle insertions and other clinical interventions. However, highly reflective parallel surfaces, such as needle walls, can create significant reverberation artifacts because the sound wave reverberates between the posterior and anterior surfaces of the object \cite{ziskin1982comet}. When the amount of reflected energy is significant, it manifests itself as an additional echo from the same surface. The reverberation artifacts are relatively bright, looking like actual boundaries which sometimes would overlap with tissue present in the image. Not only can needles and other metallic objects cause such artifacts, but certain anatomical structures with large acoustic impedance might lead to reverberation artifacts as well \cite{kirberger1995imaging}. This kind of artifact can cloud clinicians' judgement \cite{mohebali2015acoustic} and confuse medical image analysis algorithms \cite{xu2012ultrasound,reusz2014needle}. For some pixels, it can be difficult to differentiate whether the pixel is an artifact, or to assign a percentage to the pixel indicating how much of the pixel's value is artifact or actual tissue measurement. The brightness of artifacts somewhat-predictably falls off as they get further away from the reflective object, but the artifacts have uncertain boundaries and differing intensity distributions. Pixel-wise labeling is challenging and time consuming for annotators, who may have considerable differences in their annotations (example shown in Fig.~\ref{exa}). Semi-supervised approaches can reduce human labeling time, and weak supervision can estimate how much of the pixel is corrupted by the reverberation artifact. 

In this paper, we propose a novel weakly- and semi- supervised probabilistic needle-and-reverberation-artifact segmentation algorithm and show that our algorithm is useful in improving the performance of downstream tasks. The algorithm can be broken into three parts. The first part is a probabilistic segmentation network trained with hard binary and not-so-accurate labels. The probabilistic segmentation is used to cope with the labeling errors and ambiguity in the input images. The second part is a transform function that turns the output of the first network into soft-labeled images where the value of each pixel represents how much of the pixel value is affected by the artifact. The last part is another network that is trained with the newly generated labels as desired output.  The main contributions of this paper are: (1) We developed a weakly- and semi- supervised segmentation framework to segment the needle and reverberation artifacts which could cope well with the insufficiency in labeled data. (2) Our probabilistic approach is able to deal with ambiguity in the artifact boundaries and variations in human labels. (3) Our approach estimates how much of the pixel values are corrupted by reverberation artifacts.

\section{RELATED WORK}
There already exists a long history of research algorithms to identify and remove reverberation artifacts from medical images. \cite{tay2006transform} produces a near optimal estimate of the reflectivity value due to reverberation by soft
thresholding the 2D discrete wavelet transform. \cite{tay2011wavelet} improves the method proposed by \cite{tay2006transform} by utilizing radiofrequency (RF) data and soft thresholds on the wavelet coefficients to estimate the reflectivity values caused by the artifact. \cite{win2010identification} models the reverberation mathematically, identifying and removing the artifact in RF data. Temporal information is used by \cite{bylund2005interactive} when they develop the 3D filter to reduce reverberation artifacts. These methods either made use of the RF data or temporal information, i.e. video sequence, but most of the times, we won't have access to RF data or a large number of ultrasound videos.

Many deep learning segmentation frameworks for medical images can be good at segmenting reverberation artifacts too, but to the best of our knowledge, there is no learning-based segmentation framework specifically for the purpose of reverberation artifacts in ultrasound images. U-Net \cite{ronneberger2015u} uses an encoder-decoder framework with skip connections between the encoder and the decoder, enabling the network to tend to more fine-grained details. \cite{mathai2019segmentation} puts Long Short-Term Memory layers (LSTM) and U-Net together, enabling the proposed USVS-Net to excel at  identifying ambiguous boundaries and be robust against speckle noises. Attention U-Net is proposed by \cite{oktay2018attention} to suppress irrelevant regions and have the network focus more on the target structure with different shapes and sizes. \cite{roy2018concurrent} re-calibrates the channels by putting Squeeze and Excitation (SE) blocks \cite{hu2018squeeze} into a normal U-Net to increase the values of meaningful features and suppress the values of the insignificant ones. Together with dual attentive fusion blocks, adversarial learning is used in \cite{han2020semi}, allowing semi-supervision of the segmentation, instead of full supervision, so that it is able to ameliorate the problem of not having enough labeled data. These methods are robust in a fully supervised situation and are good at segmenting structures with hard labels, but they are not able to generate segmentation masks that have a meaningful probabilistic output, i.e. they are unable to model the ambiguity in the data. 

\cite{kendall2017uncertainties} explicitly models the aleatoric and epistemic uncertainty in deep learning  and increased the performance on noisy data sets with dropout layers and an additional variance term in the output. \cite{nair2020exploring} adopts Monte Carlo dropout to explore the uncertainty in the network for medical image segmentation. The work further explains how to capture different types of uncertainty in medical images. \cite{kohl2018probabilistic} 
combined a conditional variational autoencoder with a U-Net. The model does not just provide a single segmentation for a given input; instead it predicts multiple plausible segmentations drawn from the distribution learned from the training data. However, this framework only works well on images with a single object or with global variations. As a result, a hierarchical probabilistic U-Net (HPU-Net) \cite{kohl2019hierarchical} is proposed to solve such problems. Instead of global latent variables, coarse-to-fine hierarchical latent variable maps are used in this work. Besides U-Net like blocks in the decoder side, some blocks are also sampled from the prior which is trained to have similar distribution with the posterior. \cite{baumgartner2019phiseg} models the conditional probability distribution of the segmentation given an input image with a network named PHiSeg. Like the HPU-Net, PHiSeg also uses a posterior network and a prior network. However, unlike the HPU-Net, the likelihood network in this work completely samples from the prior network. \cite{meng2019weakly} proposed a weakly supervised framework to segment the shadows in ultrasound images. Essentially, the work uses one network for the initial segmentation, feeds the output of this network into a transform function, and then gets a new set of labels to train another network. The transform function in this paper transforms the hard labels to soft labels, and at the same time, reduces the effect of labeling error. 

\section{METHOD}
\subsection{Overview}
\begin{figure*}[thpb]
      \centering
      {\includegraphics[width=1.0\textwidth]{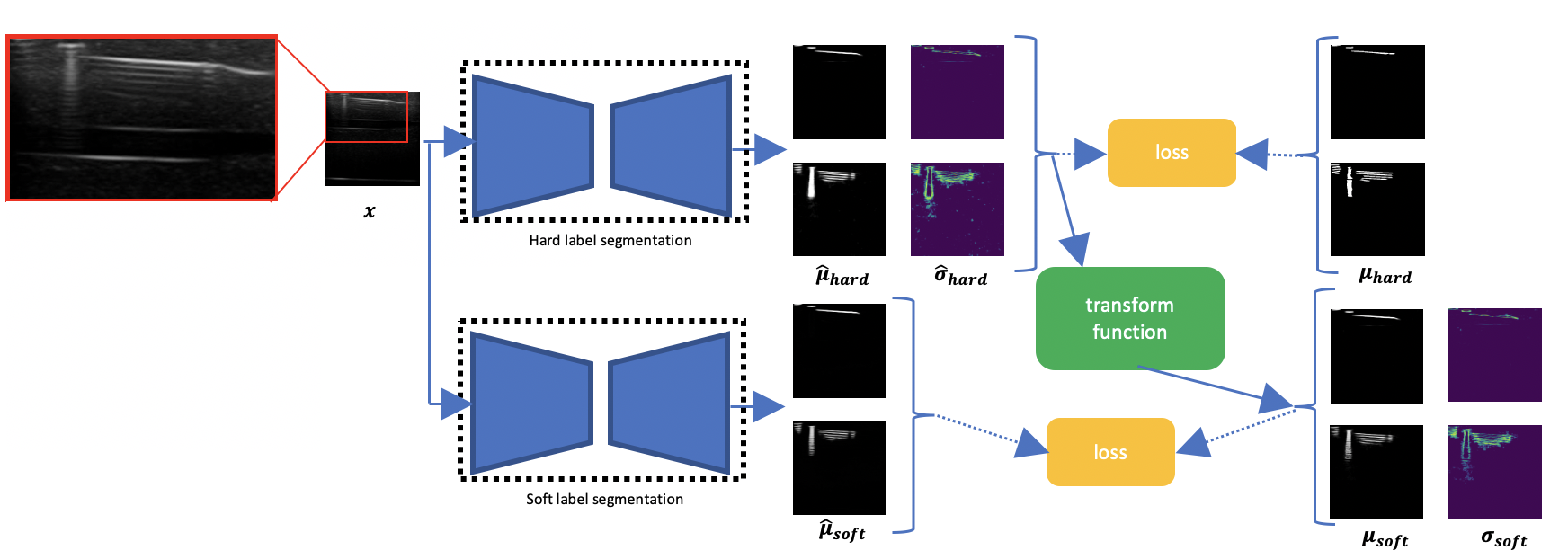}
}
      \caption{The pipeline of the algorithm is inspired by \cite{meng2019weakly}. However, due to the difference between reverberation artifacts and shadow artifacts, we use probabilistic networks and a novel transform function in our work. The pipeline can be divided into three parts: (1) probabilistic segmentation on hard labels, (2) transform function and (3) probabilistic segmentation on soft labels. We use the hard labels to train the hard label segmentation network, then transform the output of this network to soft labels by the transform function. After that we use the soft labels to train the soft label segmentation network.}
      \label{segmentation}
   \end{figure*}

Due to the difficulty in defining the actual boundaries of needles, our method is designed to deal with over-labeling from annotators. Over-labeling is a labeling scheme that includes all the pixels of interest in the label but it would falsely include background pixels in the label as well. The goal of over-labeling is to have fewer false negatives, but the tradeoff is that it also includes more false positives. Over-labeling is much faster than the traditional accurate labeling while including more true positives. Inspired by \cite{meng2019weakly}, we proposed a multi-step segmentation framework, shown in Fig.~\ref{segmentation}. The entire framework can be divided into three parts: (1) a probabilistic segmentation trained on over-labeled hard labels, (2) a transformation function which takes in the output of the first network and transforms the probabilistic outputs into soft labels to remove the false positives and quantify the artifacts, (3) another probabilistic segmentation network trained on the newly generated soft labels. Our algorithms differ from \cite{meng2019weakly} in four main parts: (1) our transform function is designed based on the appearances and physical properties of reverberation artifacts instead of shadows, (2) we not only improve the segmentation masks in the transform function, but also quantify the artifacts, (3) we propose to use probabilistic segmentation to compensate for the over-labeling by human experts instead of deterministic segmentation, (4) we propose a new probabilistic segmentation network that learns variance from a known variance posterior. Even though the first part of the algorithm is fully supervised, accurate labeling is not needed, since the probabilistic segmentation method models the ambiguity in the labels and the original inputs. After getting the first network trained, the unlabeled ultrasound images can be passed through the network, creating hard segmentations on the unlabeled images. The amount of training data for the second network can also be increased with this approach. The output of the first network is then modified by a transform function, which converts the hard labeled segmentation into soft labels. The transform function is designed specifically for the appearance and physical properties of reverberation artifacts in ultrasound images. The second network is designed to further eliminate the effects of uncertainty in human labels and ambiguity of the images, creating a soft labeled segmentation which represents how much of each pixel value is corrupted by reverberation artifacts.

\subsection{Probabilistic Segmentation on Hard Labels}
The artifacts might differ by shape, intensity distribution, and unclear boundaries, so human labels could be different across annotators. Even when the same annotator labels the same image multiple times, the results can still differ. As a result, we conjecture that careful elimination of the ambiguity in the labeling can assist with subsequent analysis. Ideally, the segmentation algorithm needs to generate nearly identical results for the same image despite using data labeled by different annotators. The work in \cite{kohl2019hierarchical} models the ambiguity in the images and the labels, which is what we look for in this stage of work. Therefore, we utilize their network for our first step: segmentation on hard labels. In our network, even though we used the same general structure as \cite{kohl2019hierarchical}, we used more local blocks than global blocks, so that we could model the ambiguity of the edges better.  

We train the network with hand labeled images, but after it is trained, we don't necessarily need to use the labeled training images for the following steps. Sampling from a learned distribution, the network will generate two mean maps and two standard deviation maps for the artifacts and needles segmentation. We denote these output mean maps for artifacts and needles as $\hat{\mu}_{artifact,hard}$ and $\hat{\mu}_{needle,hard}$ respectively. We denote the output standard deviation maps for artifacts and needles as $\hat{\sigma}_{artifact,hard}$ and $\hat{\sigma}_{needle,hard}$ respectively. 

\subsection{Transform Function}
Even though the output of the first network inherently models the ambiguity in the input images and the human labels, it does not account for the fact that weaker reverberation artifacts do not affect the quality of the image as much as the strong ones do. Therefore, we transform the segmentation mask into a soft labeling mask, which models how much the artifact affects the pixel value. There are three components to our transform function:  (1) reduction of false positives, (2) introduction of an exponential decay depending on how far the artifact is relative to the needle causing the artifact, and (3) lowering the artifact-segmentation soft labels for pixels in between adjacent reverberations.

The first part of the transform function takes in predicted hard needle mask $\hat{\mu}_{needle,hard}$ and artifact mask $\hat{\mu}_{artifact,hard}$. We begin by clustering the artifacts based on distance. We then determine if there is a needle closely above a cluster of artifacts: if there is, then keep the artifacts, and set the needle as the cause for this cluster of artifacts. This is to deal with the situation where there are multiple needles in the image. If not, then remove the cluster. The pseudo-code for the algorithm is shown in Algorithm~\ref{remove}. 
The horizontal threshold $ht$ will be small because needle artifacts are typically (near) continuous horizontal lines, whereas the vertical threshold $vt$ will be larger to encompass the vertical spacing between artifact lines, which is based on the needle's reverberating cross-section \cite{kirberger1995imaging}. The threshold $t$ indicates the largest possible distance between the segmented artifacts and the corresponding needles for the artifacts to be considered true positives. In this paper, for $256\times256$ images, we set $ht=7$, $vt=11$, and $t=10$. The hyperparameters should not affect the result much as long as they are in a reasonable range. 

\begin{algorithm}[ht]
\KwData{$\hat{\mu}_{needle,hard}$, $\hat{\mu}_{artifact,hard}$}
\KwResult{$y_{1}$:  artifact mask with false positives removed}
$B,y_0=zeros(\hat{\mu}_{artifact,hard}.shape)$; $k=1$;\\
 \For{$[\mathrm{row}\; i, \mathrm{column}\; j]$ where $\hat{\mu}_{artifact,hard}[i,j]>0$}{
 \uIf{$B[i,j]==0$}{
  create stack $s$; \\
 $s.push([i,j])$; $B[i,j]=k$;\\
 \While{$s$ is not empty}{
 $[x,y]=s.pop$;\\
 \For{$[ii,jj]$ where $(\frac{ii-x}{vt})^2+(\frac{jj-y}{ht})^2<1$}{
 \uIf{ $B[ii,jj]==0$}{
 $s.push([ii,jj])$;
 $B[ii,jj]=k$;
 }
 }
 }
 }
 $k++$;\\
 }
 $set1=where(\hat{\mu}_{needle,hard}>0.5)$;\\
 \For {$(kk=1;kk<k;kk++)$}{
 $set2=where(B==kk)$;\\
\uIf{$\exists[i_0,j_0]\in set2, [i_1,j_1]\in set1$, and $\forall [i_2,j_2]\in set1$, s.t. $i_0$ below $i_2$ and $distance([i_0,j_0],[i_1,j_1])<t$}{ $y_1+=(B==kk)*\hat{\mu}_{artifact,hard}$}


 }
 \caption{False Positives Removal}
\label{remove}
\end{algorithm}

The second part of the the transform function's purpose is to create the exponential decay in the segmentation and, at the same time, compensate for the pixels that do not comply with the decay. As sound waves travel through a medium, the intensity falls off as the distance increases. Normally, the intensity decays exponentially \cite{amin1989ultrasonic}, and due to the fact that reverberation artifacts are caused by sound waves reverberating inside the tissue \cite{ziskin1982comet}, the intensity of the reverberation artifacts should be no different. The first bounce of the sound wave at the needle represents the needle, followed by resonating bounces resulting in the artifact, so the intensity of the artifact should be on an exponential decay of the intensity of the needle. The intensity of the artifact falls until the last of the identified artifact pixels towards the bottom. We model this exponential decay as:

$$
y_{2,1}(i,j)=y_1(i,j)e^{-\alpha\frac{h(i,j)}{d(i,j)}} \eqno{(1)}
$$
where $\alpha$ is a hyper parameter, depending on the ultrasound imaging settings. The higher the frequency, the larger $\alpha$ should be, as sound waves would then encounter more resistance in-depth. For the experiments in this paper, $\alpha=0.8$. $h(i,j)$ represents the distance of pixel $(i,j)$ to the needle that is causing the artifact. $d(i,j)$ denotes the distance between the deepest pixel (the furthest pixel away from the corresponding needle) in the cluster of artifacts containing $(i,j)$ and the nearest pixel in the corresponding needle.

However, other objects and tissues in the image would also have minor effects on the pixel values in the reverberation artifacts, e.g. other boundaries which overlap with the reverberation, shadows caused by vessels interacting with the reverberations, etc. The exponential-decay artifact model does not account for these other components of the pixel values. To deal with the problem, we also create an alternate measurement $y_{2,2}$ based on the pixel values in the input images. Denote the input image as $I$. For normalization, we first find the maximum pixel value $m_1$ in the needle-region of $I$. The normalized pixel values in $I$ can then be used as weights on the artifact soft-label mask as follows:
$$
m_1=\max_{\forall(ii,jj), \hat{\mu}_{needle,hard}(ii,jj)>0.5}I(ii,jj) \eqno{(2)}
$$

$$
y_{2,2}(i,j)= \frac{I(i,j)}{m_1}y_1(i,j) \eqno{(3)}
$$
In cases where artifact pixels are unusually bright, e.g. $y_{2,2}$ is large due to overlapping with actual object boundaries. Preserving such property is desired because it represents the actual anatomy or a different artifact (such as a diagnostic B-line in a lung).  We therefore combine $y_{2,1}$ and $y_{2,2}$ by taking the maximum

$$
y_{2}(i,j)= \max(y_{2,1}(i,j),y_{2,2}(i,j)) \eqno{(4)}
$$

At this point, we have removed the false positives and also created an exponential decay in the artifact soft labels while preserving the effects of underlying anatomic boundaries on artifacts. However, we still need to assign lower values to the soft labels in between each reverberation, since the hard human labels tend to (incorrectly) in-paint such regions. The pixels between reverberations tend to have lower values than the artifact pixels in the original input images. In the third part of our transform function we determine whether a certain pixel belongs to the non-artifact region in between reverberation lines.  We need to compare the value of that pixel in the original input ultrasound image against the maximum value in a local patch. We also apply a sigmoid-like function to make sure that artifact soft labels are limited with respect to the highest intensity value within that region.

$$
m_2(i,j)=\max_{(ii,jj)\in \Omega_1(vw,hw)}I(i+ii,j+jj) \eqno{(5)}
$$

$$
\mu_{artifact,soft}(i,j)= \frac{y_2(i,j)}{1+e^{-\frac{\beta I(i,j)}{m_2(i,j)}+\frac{\beta}{2}}} \eqno{(6)}
$$
where $\beta$ is a hyper-parameter that controls how fast the fall-off is. If the noise level is high, a larger $\beta$ should be used. Also. $\Omega_1(vw,hw)$ is a rectangular region where $(0,0)$ is the center point, $2vw$
and $2hw$ is the height and width. $vw$ and $hw$ stand for vertical and horizontal window respectively. $vw$ should be large enough to include at least one line of true reverberation artifact in the patch.  In this paper, we set $\beta=8$, $vw=2$, and $hw=1$. Again, these parameters are not sensitive, as long as $vw$ and $hw$ stay relatively small.

As for the standard deviation map of artifacts, we want to rescale its values in the same manner as we rescaled the mean map of artifacts. Therefore, the transform function for the standard deviation map can be simplified to 
$$
\sigma_{artifact,soft}(i,j)= \frac{\mu_{artifact,soft}(i,j)}{\hat{\mu}_{artifact,hard}(i,j)+\epsilon} \eqno{(7)}
$$
where $\epsilon \ll 1$ avoids division by zero.

The needle labels will not change in this part, as a result, we have $\mu_{needle,soft}(i,j)=\hat{\mu}_{needle,hard}(i,j)$, and $\sigma_{needle,soft}(i,j)=\hat{\sigma}_{needle,hard}(i,j)$.

\subsection{Probabilistic Segmentation on Soft Labels}

\begin{figure*}[thpb]
      \centering
      {\includegraphics[width=1.0\textwidth]{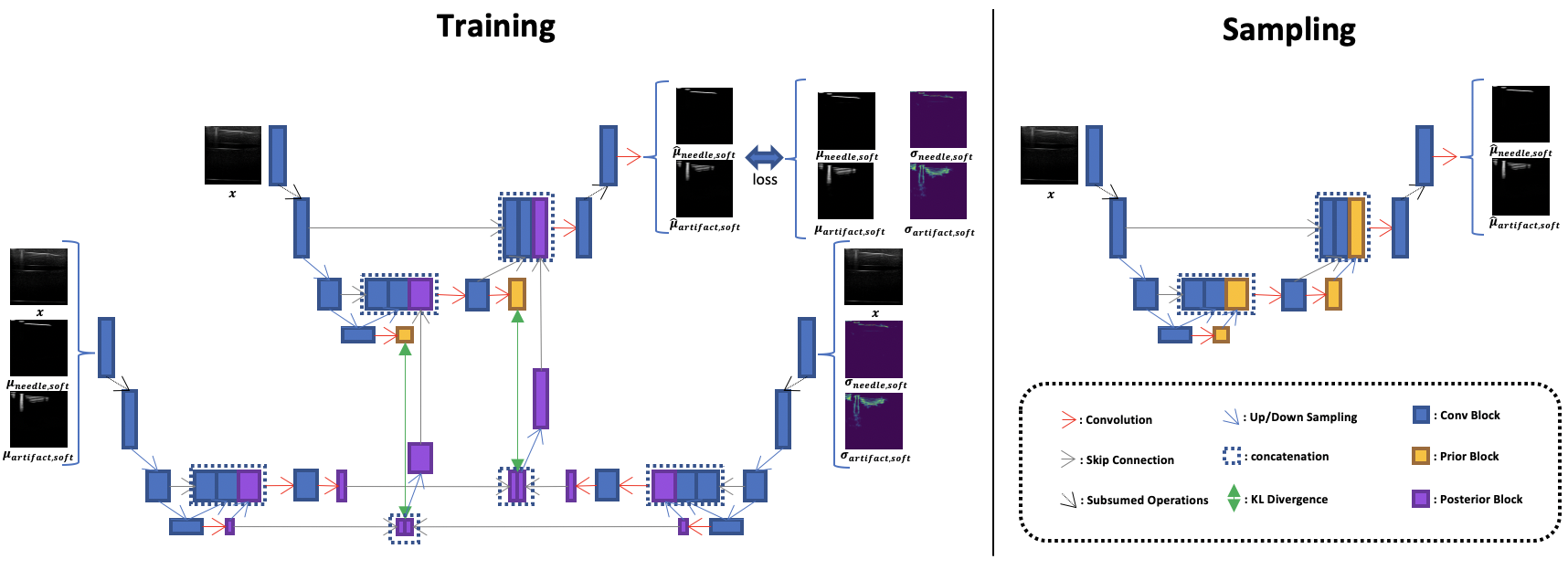}
}
      \caption{Our proposed probabilistic network is similar to Hierarchical Probabilistic U-Net \cite{kohl2019hierarchical}, but instead of learning the posterior with labeled segmentation masks, our method learns mean and variance of the posterior from known mean and variance separately. Therefore, two variational autoencoders are used in the posterior.}
      \label{network}
   \end{figure*}

Our segmentation model here is built upon the HPU-Net \cite{kohl2019hierarchical}, and we follow their notation below. We want to model the unknown ambiguity in the images and the labels, and at the same time, we also want to take the known variance in the labels into account. Our training objective is similar to that of the previous work: maximizing the evidence lower bound on the likelihood $p(M|X)$, except that we are modeling a variational posterior $Q(.|X,M,V)$ instead of $Q(.|X,M)$, where $X$ is the input image, $M$ is the known mean of the segmentation label, and $V$ is the variance of the segmentation label. We calculate the posterior $Q$ from two separate networks, where one network accounts for the mean $\mu_i^{post}(z_{<i},X,M)$ and the other network models the variance $\sigma_i^{post}(z_{<i},X,V)$. The latent features in the prior blocks should follow a normal distribution generated by the posterior blocks $N(\mu_i^{post}(z_{<i},X,M),\sigma_i^{post}(z_{<i},X,V))$. During training, we directly sample from the posterior $Q$, and train the normal distribution generated by the prior $N(\mu_i^{prior}(z_{<i,X},\sigma_i^{prior}(z_{<i,X}))$ to be close to the one from the posterior. When sampling, we did exactly the same as the previous work did: sampling the latent features from the normal distribution modeled by the prior. The training and sampling process is illustrated in Fig.~\ref{network}. 

In this particular application, we utilize a mean-squared-error-based custom loss function as a way to deal with the continuous values and unique meaning of soft labels. To deal with overfitting to the background, we only care about the pixels that have values over a certain threshold $\gamma$. Lower weights are assigned to pixels where absolute error is within the known standard deviation, since we are less sure about the value in the label where standard deviation is larger. Therefore, the loss function can be expressed as the following
$$
loss=\sum_{\forall(i,j)\in \Omega_2}w(i,j)(\hat{\mu}_{soft}(i,j)-\mu_{soft}(i,j))^2 \eqno{(8)}
$$
where $$\Omega_2=\{(x,y)| \hat{\mu}_{soft}(x,y)>\gamma \lor \mu_{soft}(x,y)>\gamma \} \eqno{(9)}$$ 
$$ w(i,j)=\left\{
\begin{aligned}
k \ abs(i,j)<\sigma_{soft}(i,j)\\
1 \ abs(i,j)\geq \sigma_{soft}(i,j) \\
\end{aligned}
\right. \eqno{(10)}
$$
$$ 
abs(i,j)=|\hat{\mu}_{soft}(i,j)-\mu_{soft}(i,j)| \eqno{(11)}
$$
and $k<1$.

\section{EXPERIMENTS}
\label{exp}
\begin{figure}[thpb]
      \centering
      \subfloat[]{\includegraphics[width=0.15\textwidth]{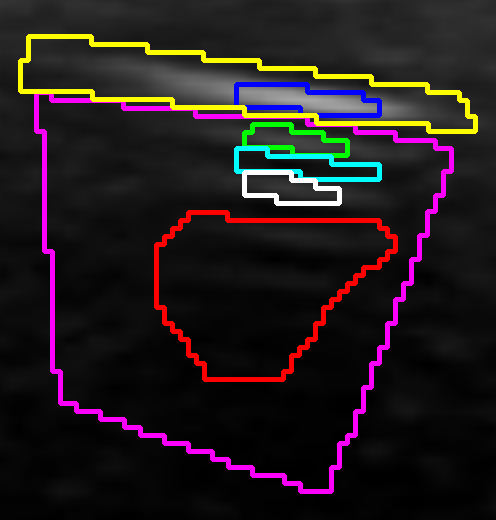}

      \label{label}
      }
      \subfloat[]{\includegraphics[width=0.15\textwidth]{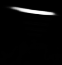}

      \label{needle}
      }
      \subfloat[]{\includegraphics[width=0.15\textwidth]{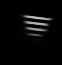}

      \label{artifact}
      }
      
      \caption{Labels and segmentation results. In Fig.~\ref{label}, yellow and pink labeled areas represent the possible areas containing needles and artifacts respectively, green and white labeled regions indicate the patches that are definitely the first and second reverberations, the bright blue circle(s) represent the non-artifact patches between reverberations, the dark blue region(s) are the most confident needle pixels, and the red patches are the regions with identifiable fuzzy artifacts. }
   \end{figure}

 The ultrasound imaging is performed with a UF-760AG Fukuda Denshi diagnostic ultrasound imaging machine, on both a chicken breast and an anthropomorphic phantom containing simulated vessels in a tissue medium produced by Advanced Medical Technologies. We used a linear transducer with 51mm scanning width. It was set to 12MHz with an imaging depth of 4cm and a gain value of 21dB. The dimension of each image is $512 \times532$ pixels and the pixel pitch is 0.10mm horizontally and 0.075mm vertically. The images were then resized to $256\times 256$. Our models are trained on a single Nvidia Titan RTX for 15 epochs with horizontal flipping, Gaussian blurring and Gamma transform as data augmentation. There are 401, 100 and 40 images in the training, validation and test set respectively. Our dataset is split in this ratio due to the sophisticated labeling of the test set. The labeling in this work utilized the labeling tool Labelme \cite{labelme2016}. The training set is over-labeled to show that our algorithm is good at dealing with inaccurate labels as well as with handling false positives in the training data.  A example of our test set labeling is shown in Fig.~\ref{label}, where the yellow and pink regions indicate the possible areas containing needles and artifacts, respectively, green and white regions represent the first and second reverberations that we are confident in, the bright blue circles represent the patch between reverberation with no artifacts, the dark blue represents the needle patches that we are confident in, and the red patches are the regions with identifiable fuzzy artifacts. We define the artifact and needle pixels with values greater than 0.05 and 0.5 as positive, respectively. The labeling is designed to show that in our result: (1) the pixels that definitely belong to needles and to the significant artifacts are segmented, (2) the pixels that are between the reverberations and not part of the artifacts are not segmented, (3) the fuzzy artifacts are segmented with lower values, (4) very few false positives are included. 
 
 Our notations are summarized in Table~\ref{notation}. We denote the ratio between positive artifact and needle counts in the labeled first, second reverberations and needle and the total pixels counts in those regions as FAR (first artifact rate), SAR (second artifact rate), NR (needle rate) respectively, and the average value for positives in those artifact patches as FAA (first artifact average) and SAA (second artifact average). We also denote the false positive rate of artifacts and needles as AFPR (artifact false positive rate) and NFPR (needle false positive rate), and the average value of the false positives as AFPA (artifact false positive average) and NFPA (needle false positive average). We also calculate the average of the non-artifact region between reverberations as NAA (non-artifact average) We further define the average positive artifact value in the identifiable fuzzy artifact patches as IFAA (identifiable fuzzy artifact average).

\begin{table}[h]
\caption{Notation and Abbreviations}
\label{notation}
\begin{center}
\begin{tabularx}{0.45\textwidth}{|l|X|}
\hline
FAR&ratio between positive artifact pixel count in the first labeled reverberation patch and the total pixel count in the patch\\
\hline
SAR&ratio between positive artifact pixel count in the second labeled reverberation patch and the total pixel count in the patch\\
\hline
NR&ratio between needle pixel count in the labeled needle patch and the total pixel count in the patch\\
\hline
FAA&average output value in the first labeled reverberation patch\\
\hline
SAA&average output value in the second labeled reverberation patch\\
\hline
AFPR&false positive rate of artifacts\\
\hline
NFPR&false positive rate of needles\\
\hline
AFPA&average output value of false positive artifacts\\
\hline
NFPA&average output value of false positive needles\\
\hline
NAA&average output value of non-artifact patch between reverberations\\
\hline
IFAA&average output value of identifiable fuzzy artifacts\\
\hline

\end{tabularx}
\end{center}
\end{table}
   
\begin{figure}[thpb]
      \centering
      {\includegraphics[width=0.48\textwidth]{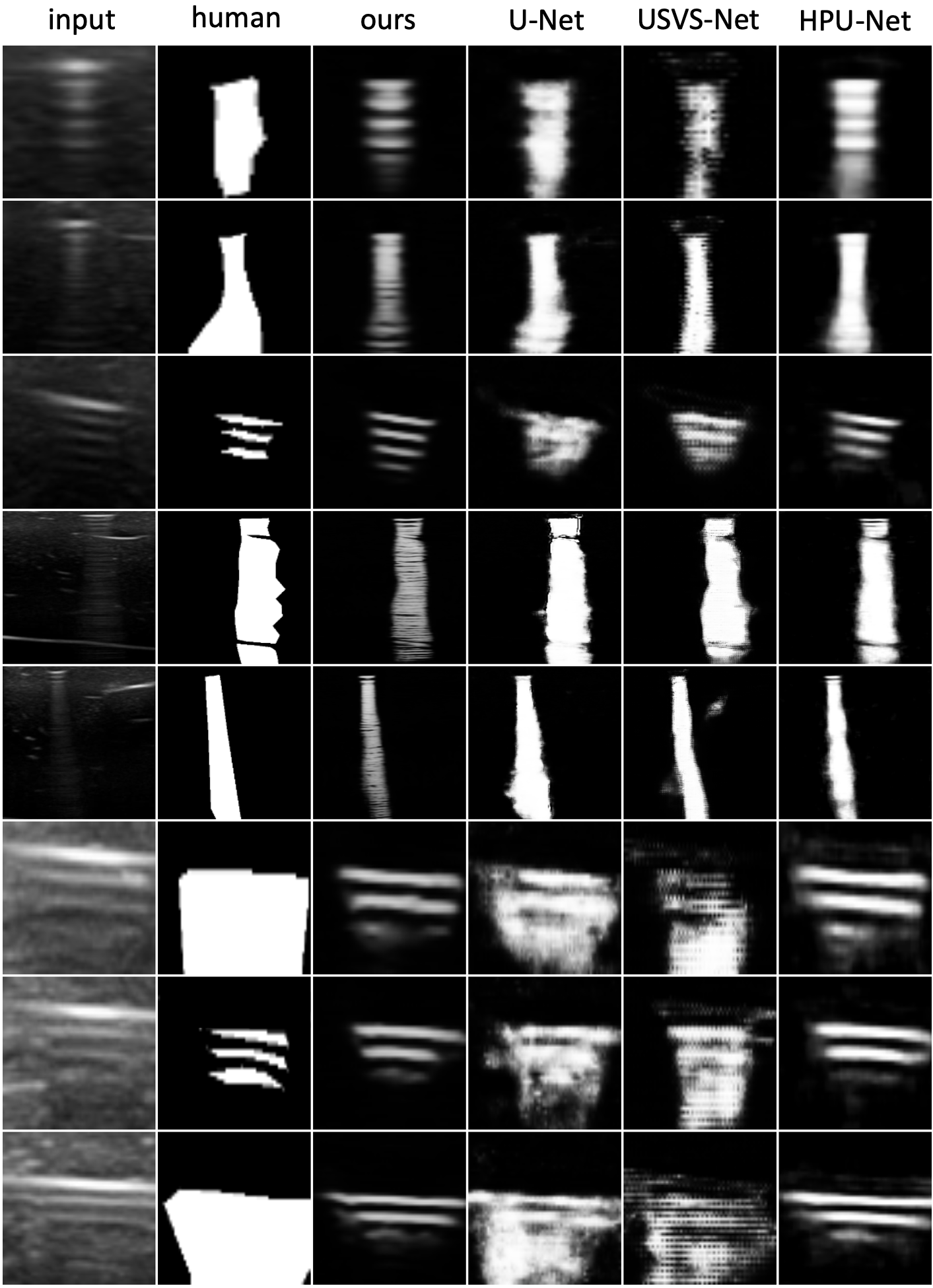}
}
      \caption{Results comparison: the first five rows are images on phantom data, last three rows are images on chicken breast data. From left to right: Input images (zoomed in), Human Labels, Our results, U-Net, USVS-Net, HPU-Net. Notice that our method differentiates the reverberations better, produces fewer false positives, picks up unlabeled artifacts and at the same time, models the exponential decay of artifact intensity. }
      \label{ex1}
   \end{figure}

In our first experiment, we compared our results against those from expert labels, U-Net \cite{ronneberger2015u}, USVS-Net \cite{mathai2019segmentation}, and HPU-Net \cite{kohl2019hierarchical} to show that our proposed algorithm is superior to these current segmentation networks and expert labels for needle-and-reverberation-artifact segmentation. The experts tend to label the entire region that might contain artifacts, but they do not differentiate between the reverberations. The quantitative results can be found in Table~\ref{table1}. As can be seen in the table, our approach performs similarly to HPU-Net at binary-thresholded segmentation of the significant artifacts and needles, and our approach performs better than the other two methods. Our goal, however, is more than just binary segmentation.  Since the value in the output represents how much the artifact affects the underlying ``real'' pixel values, and there should be a decay in the intensity of the artifacts, the average value of first artifacts should be larger than that of second artifacts. Together, they also shouldn't be too close to 1 as they normally do not completely obscure the real tissue there. \emph{Our method is the only algorithm that has achieved such results.} 
Our algorithm also limits the false positive rate to around 0.02 whereas the best results achieved by other algorithms are 0.084 and 0.04, for artifacts and needles, respectively. Our algorithm also does a good job in differentiating between reverberations, since the average value in non-artifact patches between reverberations is 0.042, which is the lowest by far, compared to 0.506, 0.431, and 0.136 from U-Net, USVS-Net and HPU-Net respectively. The IFFA value further indicates that our network learns the decay in the artifact intensity. Sample results of needle segmentation and artifact segmentation can be found in Fig.~\ref{needle} and Fig.~\ref{artifact}. Qualitative results of artifact segmentation are shown in Fig.~\ref{ex1}. Our method clearly does the best in differentiating between the actual artifacts and the patches around them, while simultaneously assigning lower values to less significant artifacts. 
   
\begin{table}[h]
\caption{Comparison of our Entire Approach Against Other Algorithms and Human Labels}
\label{table1}
\begin{center}
\begin{tabular}{|c c c c c c|}
\hline
 & Human Labels & Ours & U-Net & USVS-Net & HPU-Net\\
\hline
FAR&0.966&0.986&0.978&0.859&\textbf{0.996}\\
SAR&0.993&0.994&\textbf{1.0}&0.976&\textbf{1.0}\\
NR&0.939&\textbf{0.981}&0.854&0.786&0.966\\
FAA&0.922&\textbf{0.728}&0.806&0.589&0.894\\
SAA&0.975&\textbf{0.698}&0.928&0.831&0.936\\
AFPR&0.129&\textbf{0.022}&0.237&0.168&0.084\\
NFPR&0.119&\textbf{0.018}&0.058&0.329&0.040\\
AFPA&0.934&\textbf{0.151}&0.352&0.261&0.130\\
NFPA&0.914&\textbf{0.627}&0.715&0.750&0.716\\
NAA&0.423&\textbf{0.042}&0.506&0.431&0.136\\
IFAA&0.990&\textbf{0.471}&0.871&0.836&0.775\\
\hline
\end{tabular}
\end{center}
\end{table}

Our second experiment is done to show that our proposed second network in the pipeline works better than other networks. In this experiment, we test out U-Net, HPU-Net, and USVS-Net as the second network. We found that USVS-Net overfits to the background too much, so it is not included in the quantitative comparison. The quantitative evaluation can be found in Table~\ref{table2}. It shows that our method is as good as HPU-Net in detecting the strong artifacts and needles, since the FAR, SAR and NR values are similar. It also illustrates that our method works better in differentiating between the artifacts and the patches between reverberations, as well as giving higher values to the first reverberation than to the second one. Lastly, our methods also produce fewer false positives and has lower values in the false-positive patches.

\begin{table}[h]
\caption{Comparison of our Second Network Against Other Soft-Label Probabilistic Networks}
\label{table2}
\begin{center}
\begin{tabular}{|c c c c|}
\hline
 & Ours & U-Net & HPU-Net\\
\hline
FAR&\textbf{0.986}&0.978&\textbf{0.986}\\
SAR&0.994&\textbf{1.0}&\textbf{1.0}\\
NR&0.981&0.855&\textbf{0.990}\\
FAA&\textbf{0.728}&0.806&0.685\\
SAA&\textbf{0.698}&0.928&0.686\\
AFPR&\textbf{0.022}&0.237&0.060\\
NFPR&\textbf{0.018}&0.058&0.066\\
AFPA&\textbf{0.151}&0.352&\textbf{0.151}\\
NFPA&\textbf{0.627}&0.715&0.725\\
NAA&\textbf{0.042}&0.506&0.064\\
IFAA&\textbf{0.471}&0.871&0.476\\
\hline
\end{tabular}
\end{center}
\end{table}

Our next experiment is to show that we can further improve our results with additional unlabeled data, and that if the training data of the first network is unavailable, using only non-labeled data also wouldn't hurt the performance much. The comparison can be found in Table~\ref{table3}, where it shows that using both labeled and unlabeled data slightly decreases the false positives of needles, and also lowers the values between reverberations. On the other hand, using only unlabeled data produces only slightly more false positives.

\begin{table}[h]
\caption{Benefit from Adding Labeled or Unlabeled Data}
\label{table3}
\begin{center}
\begin{tabular}{|c c c c|}
\hline
 & labeled only & both & unlabeled only\\
\hline
FAR&0.986&\textbf{0.988}&0.971\\
SAR&0.994&\textbf{0.999}&0.996\\
NR&\textbf{0.981}&0.966&0.941\\
FAA&\textbf{0.728}&0.723&0.698\\
SAA&\textbf{0.698}&0.695&0.691\\
AFPR&\textbf{0.022}&0.029&0.043\\
NFPR&0.018&\textbf{0.005}&0.024\\
AFPA&0.151&0.143&\textbf{0.125}\\
NFPA&0.627&\textbf{0.624}&0.643\\
NAA&0.042&\textbf{0.033}&0.040\\
IFAA&0.471&\textbf{0.440}&0.497\\
\hline

\end{tabular}
\end{center}
\end{table}

\begin{figure}[thpb]
      \centering
      {\includegraphics[width=0.45\textwidth]{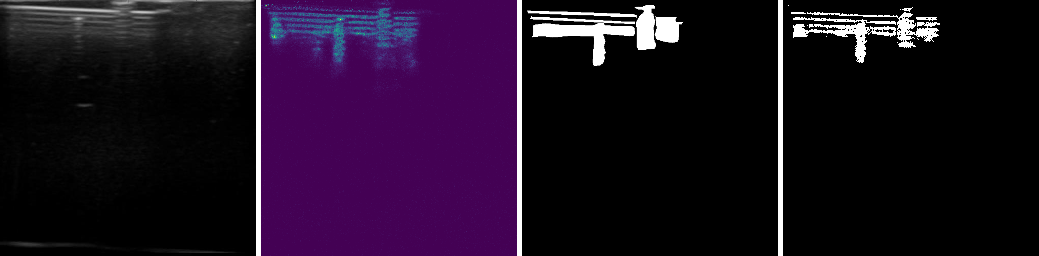}
}
      \caption{Example of aleatoric uncertainty map and uncertainty-pruned label. Left to right: input image, aleatoric uncertainty map, hand label, uncertainty-pruned label}
      \label{refined}
   \end{figure}

To provide further validation of our algorithm on different sets of plausible hand labels, we apply aleatoric uncertainty maps obtained from a Monte Carlo dropout-based Bayesian 3-D convolutional neural network similar to \cite{yongchankwon2018uncertainty}. In doing so, we are able to simulate the wave-like intensity decrease of needle artifacts in a realistic manner. For modeling the aleatoric uncertainty in the artifact segmentation region, we apply the model estimator as
$$
    \frac{1}{T} \sum_{t=1}^{T} diag(\hat{p}_t) - \hat{p}_t^{{\bigotimes}2} \eqno{(12)}
$$
using the notation from \cite{yongchankwon2018uncertainty}, where  $\hat{p}_t = p(\hat{w}_t)$ represents softmax outputs evaluated with a set of random weights $w$,In an attempt to obtain more articulate hand labels, we apply the following steps on top of our hand labels: (1) use the labels to mask only the relevant regions in the aleatoric uncertainty map, (2) take the inverse of the uncertainty map, and (3) determine a local threshold $a_{thresh}$ for smaller patches within each labeled region and use that for deciding whether to keep or remove the labels. A sample of the aleatoric uncertainty map and uncertainty-pruned label are shown in Fig~\ref{refined}. We compare the results from networks trained on the original hand labels against the ones trained on the refined labels. The results are shown in Table~\ref{table4}. Our algorithm still does a decent job on the uncertainty-pruned labels, even though the performance is slightly worse than on original hand labels. The intuition is that although the uncertainty-pruning removes some of the labels in the spaces between reverberations. Our algorithm is designed to differentiate the reverberations from the spaces between reverberations to compensate for the over-labeling, so it does not make much difference in that regards. However, the pruning step removes some of the true positives as well, which could not be learned in the networks, slighting hurting the performance. The algorithm still does a decent job on a different set of labels, while performing slightly better on the over-labeled annotations.

\begin{table}[h]
\caption{Our Robustness Across Various Binary Labels}
\label{table4}
\begin{center}
\begin{tabular}{|c c c|}
\hline
 & hand labels & uncertainty-pruned labels\\
\hline
FAR&\textbf{0.986}&0.958\\
SAR&0.994&\textbf{0.997}\\
NR&\textbf{0.981}&0.955\\
FAA&\textbf{0.728}&0.630\\
SAA&\textbf{0.698}&0.668\\
AFPR&\textbf{0.022}&0.063\\
NFPR&\textbf{0.018}&0.034\\
AFPA&\textbf{0.151}&0.137\\
NFPA&0.627&\textbf{0.621}\\
NAA&\textbf{0.042}&0.044\\
IFAA&0.471&\textbf{0.400}\\
\hline
\end{tabular}
\end{center}
\end{table}

\section{APPLICATIONS}
In this section, we show that our reverberation artifact segmentation and quantification method is useful in several downstream tasks, such as multi-view image compounding and vessel segmentation. All of the imaging parameters are set exactly as they are in Section \ref{exp}. 
\subsection{Multi-view Image Compounding}
The goal of multi-view image compounding is to take the information from images taken at different viewpoints, and reconstruct the true underlying structure. This is task is extremely important in ultrasound, since ultrasound images are very path-dependent, thus certain structures can be seen in an image taken from one viewpoint but cannot be seen in an image from a different viewpoint. However, the same object can cast reverberation artifacts in different directions in images from different viewpoints, making multi-view compounding a challenging task. In compounding, we want to preserve the real objects and structures in the compounded image but remove the artifacts at the same time.

We apply our reverberation artifact segmentation and quantification algorithm in image compounding task, to remove the reverberation artifacts in the compounded image. We propose to use a simple compounding algorithm that considers our reverberation artifact segmentation and quantification. For example, we can take two viewpoints - even though the simple algorithm can be easily expanded to more than two viewpoints. Denote the image from viewpoint $k$ as $I_k$, our soft segmentation mask for image $I_k$ as $M_k$, the compounded image as $\hat{I}$. We define a confidence map $C$ that depicts how much of the pixel is not corrupted artifact, thus $C=1-M$. For every pixel $(i,j)$, if $C_1(i,j)-C_2(i,j)>t$, then we set $\hat{I}(i,j)=I_1(i,j)$; if $C_2(i,j)-C_1(i,j)>t$, then we set $\hat{I}(i,j)=I_2(i,j)$; else, we set $\hat{I}(i,j)=\max(I_1(i,j),I_2(i,j))$, where $t$ is a confidence threshold which we set as $0.1$ in this paper. In other words, if the confidence from one viewpoint is significantly higher, we take the pixel value from that viewpoint. Otherwise, we take the maximum image intensity across different viewpoints.

\begin{figure*}[thpb]
      \centering
      {\includegraphics[width=0.98\textwidth]{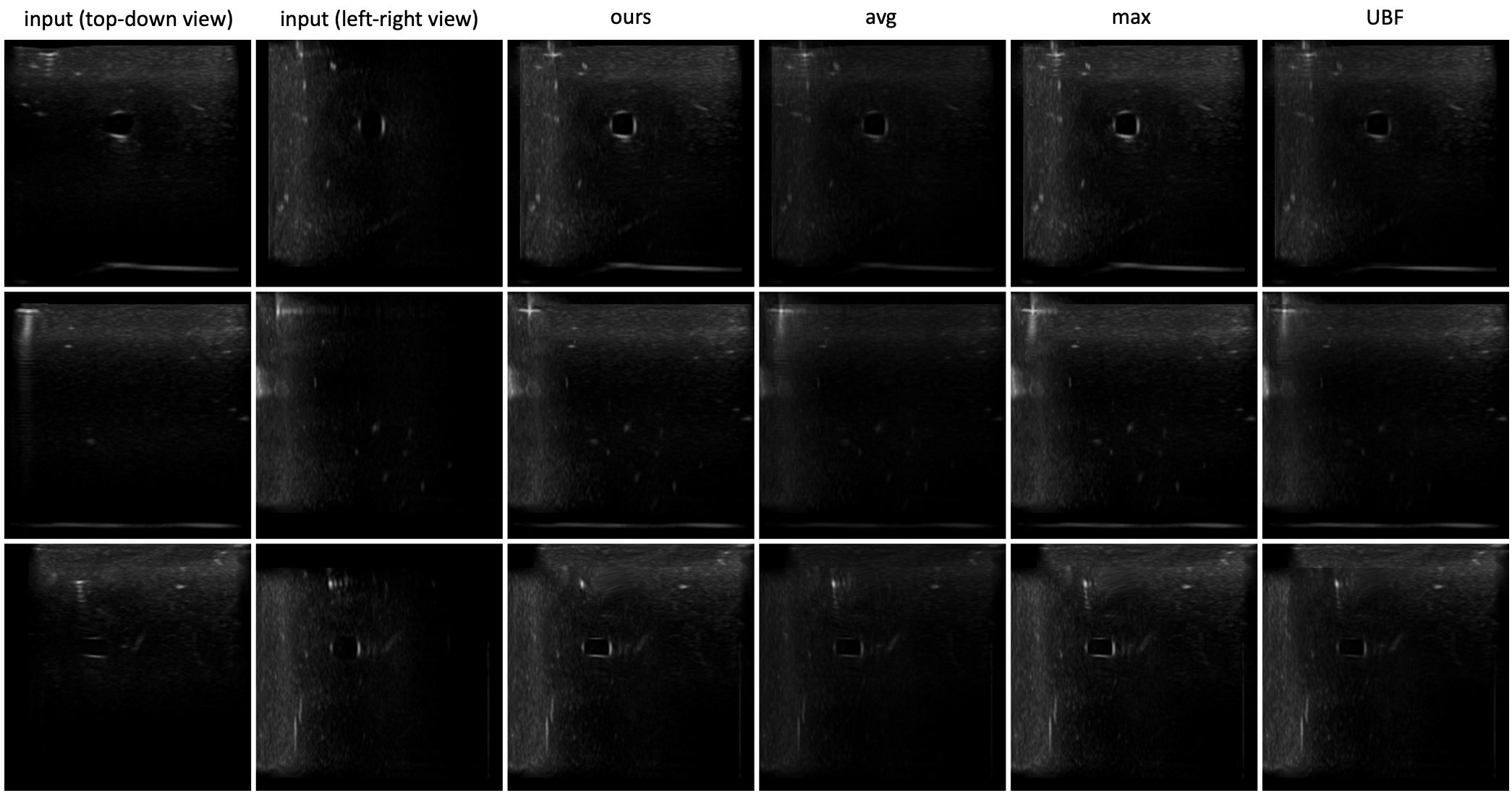}
}
      \caption{Compounding results. Left to right: two images from orthogonal view points, results from our method, average \cite{trobaugh1994frameless}, maximum \cite{lasso2014plus}, uncertainty-based fusion \cite{zu2014orientation}. It shows that our method preserves all the structures while removing the reverberation artifacts.}
      \label{compound}
   \end{figure*}

As an experiment, we take pairs of ultrasound images from orthogonal viewpoints with a needle inserted into the phantom. Since we are performing the experiment on a square phantom, it is easy to make sure the viewpoints are orthogonal with free-hand imaging. We compare our simple method against existing compounding algorithms, such as average (avg) \cite{trobaugh1994frameless}, maximum (max) \cite{lasso2014plus}, and uncertainty-based fusion (UBF) \cite{zu2014orientation}. The results are shown in Fig.\ref{compound}. We can observe that our simple compounding method preserves the vessels and needles as good as max \cite{lasso2014plus} while nearly removing all the reverberation artifacts. 

\subsection{Vessel Segmentation}
Ultrasound is often used to monitor robot-controlled surgeries, so ultrasound vessel segmentation is important for diagnostic purposes and needle insertion by robotic systems \cite{guerrero2007real}. However, needle reverberation artifacts could occlude the structures or objects of interest \cite{mohebali2015acoustic}. Such artifacts could also affect the performance of vessel segmentation \cite{reusz2014needle} and needle tracking algorithms \cite{xu2012ultrasound}. 

We believe our reverberation artifact segmentation and quantification algorithm could help with vessel segmentation when reverberation artifacts produced by needles are present in the image. During convolution, segmentation networks should pay less attention to the artifact pixels, so we propose to use the soft segmentation results from our algorithm as the masks in partial convolution \cite{liu2018image}. Although the partial convolution method is built for image inpainting, the idea to mask certain regions during convolution is useful for our purpose, since we do not want the network to treat the artifact pixels equally as other pixels. 

In this work, we use U-Net \cite{ronneberger2015u} as the backbone for the vessel segmentation. We compare our artifact-segmentation-and-quantification-based partial convolutional U-Net (partial U-Net) against a normal U-Net. For fair comparison, we have 5 layers in the encoder side, with 32, 64, 128, 256, 512 filters for both the partial U-Net and the normal U-Net, and the decoder side is exactly symmetric to the encoder. There are 481 images in the training set, and 97 images in both validation and test set. The only augmentation technique used during training is horizontal flipping. 

\begin{figure}[h!]
      \centering
      {\includegraphics[width=0.48\textwidth]{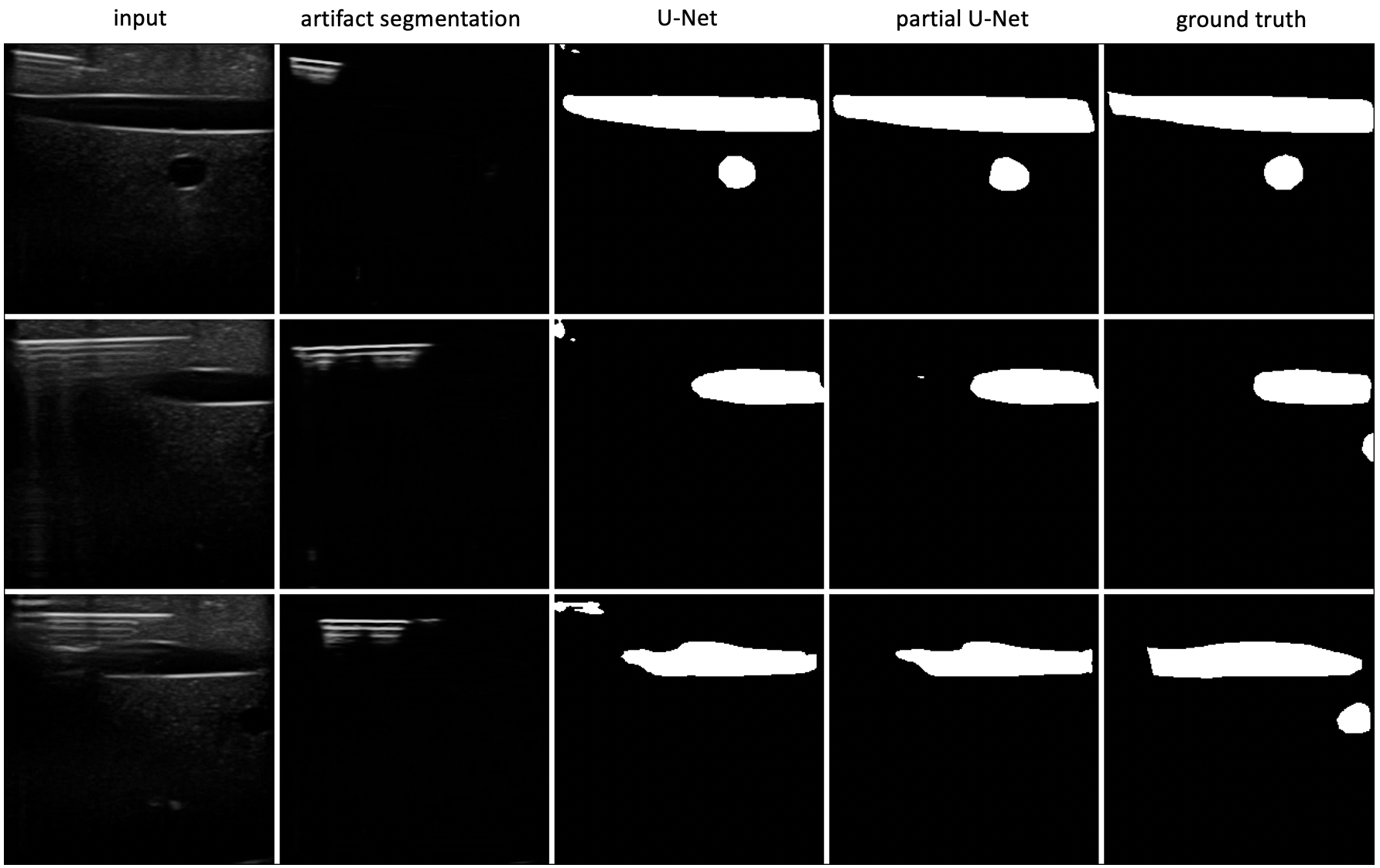}
}
      \caption{Left to right: input images, reverberation artifact masks, results from U-Net, results from partial U-Net, and ground truths. The partial U-Net uses the reverberation artifact masks to mask the convolution and does not generate false positive where needles are present like U-Net.}
      \label{partial}
   \end{figure}

Shown in the qualitative results in Fig.~\ref{partial}, the results by normal U-Net creates some false positives at the top left corner, where the needles are present. The needles together with the image boundaries or some other artifacts confuse the normal U-Net, since they look like vessel boundaries to some extent, while the partial U-Net does not generate false positive there. Quantitative results can be found in Table~\ref{table5}. The precision for artifact-masked partial U-Net is notably higher than normal U-Net, indicating that our reverberation artifact quantification helps reduce the false positives in vessel segmentation. By other metrics, the partial U-Net also outperforms normal U-Net with the help of our soft reverberation artifact segmentation. 
   
   \begin{table}[h]
\caption{Comparison Between Partial U-Net and U-Net}
\label{table5}
\begin{center}
\begin{tabular}{|c|c c c c|}
\hline
 & Precision & Recall&IoU&Dice\\
\hline
U-Net&0.823&0.801&0.684&0.812\\
partial U-Net&\textbf{0.847}&\textbf{0.803}&\textbf{0.701}&\textbf{0.825}\\
\hline
\end{tabular}
\end{center}
\end{table}

\section{CONCLUSION}
Our proposed algorithm outperforms other existing segmentation algorithms in segmenting and quantifying reverberation artifacts. Our weakly- and semi-supervised reverberation segmentation framework is able to take as input inaccurate human labels and predict accurate segmentation maps as well as quantitatively estimate how much the artifacts affect the pixel values. It works even better on over-labeled annotations. Our quantification of reverberation artifacts is useful in image compounding, and for improving vessel segmentation results. Our probabilistic outputs could also potentially be used in ultrasound image uncertainty measurements, ultrasound image quality evaluation, and reverberation artifacts removal. Lastly, this algorithm is also directly related to robot-controlled needle insertion, in which our work can be used to guide where to insert the needle, monitor the tissue under the needle, and determine the quality of the image.

\section*{ACKNOWLEDGMENT}

We would like to thank our collaborators at the University of Pittsburgh, Triton Microsystems, Inc., Accipiter Systems, Inc., Sonivate Medical, Inc., and URSUS Medical, LLC.

\bibliographystyle{IEEEtran}
\bibliography{IEEEabrv,IEEEexample}

\begin{thebibliography}{10}
\providecommand{\url}[1]{#1}
\csname url@rmstyle\endcsname
\providecommand{\newblock}{\relax}
\providecommand{\bibinfo}[2]{#2}
\providecommand\BIBentrySTDinterwordspacing{\spaceskip=0pt\relax}
\providecommand\BIBentryALTinterwordstretchfactor{4}
\providecommand\BIBentryALTinterwordspacing{\spaceskip=\fontdimen2\font plus
\BIBentryALTinterwordstretchfactor\fontdimen3\font minus
  \fontdimen4\font\relax}
\providecommand\BIBforeignlanguage[2]{{%
\expandafter\ifx\csname l@#1\endcsname\relax
\typeout{** WARNING: IEEEtran.bst: No hyphenation pattern has been}%
\typeout{** loaded for the language `#1'. Using the pattern for}%
\typeout{** the default language instead.}%
\else
\language=\csname l@#1\endcsname
\fi
#2}}

\bibitem{ziskin1982comet}
M.~Ziskin, D.~Thickman, N.~Goldenberg, M.~Lapayowker, and J.~Becker, ``The
  comet tail artifact.'' \emph{Journal of Ultrasound in Medicine}, vol.~1,
  no.~1, pp. 1--7, 1982.

\bibitem{kirberger1995imaging}
R.~M. Kirberger, ``Imaging artifacts in diagnostic ultrasound—a review,''
  \emph{Veterinary Radiology \& Ultrasound}, vol.~36, no.~4, pp. 297--306,
  1995.

\bibitem{mohebali2015acoustic}
J.~Mohebali, V.~I. Patel, J.~M. Romero, K.~M. Hannon, M.~R. Jaff, R.~P.
  Cambria, and G.~M. LaMuraglia, ``Acoustic shadowing impairs accurate
  characterization of stenosis in carotid ultrasound examinations,''
  \emph{Journal of vascular surgery}, vol.~62, no.~5, pp. 1236--1244, 2015.

\bibitem{xu2012ultrasound}
X.~Xu, Y.~Zhou, X.~Cheng, E.~Song, and G.~Li, ``Ultrasound intima--media
  segmentation using hough transform and dual snake model,'' \emph{Computerized
  Medical Imaging and Graphics}, vol.~36, no.~3, pp. 248--258, 2012.

\bibitem{reusz2014needle}
G.~Reusz, P.~Sarkany, J.~Gal, and A.~Csomos, ``Needle-related ultrasound
  artifacts and their importance in anaesthetic practice,'' \emph{British
  journal of anaesthesia}, vol. 112, no.~5, pp. 794--802, 2014.

\bibitem{tay2006transform}
P.~C. Tay, S.~T. Acton, and J.~Hossack, ``A transform method to remove
  ultrasound artifacts,'' in \emph{2006 IEEE Southwest Symposium on Image
  Analysis and Interpretation}.\hskip 1em plus 0.5em minus 0.4em\relax IEEE,
  2006, pp. 110--114.

\bibitem{tay2011wavelet}
P.~C. Tay, S.~T. Acton, and J.~A. Hossack, ``A wavelet thresholding method to
  reduce ultrasound artifacts,'' \emph{Computerized Medical Imaging and
  Graphics}, vol.~35, no.~1, pp. 42--50, 2011.

\bibitem{win2010identification}
K.~K. Win, J.~Wang, C.~Zhang, and R.~Yang, ``Identification and removal of
  reverberation in ultrasound imaging,'' in \emph{2010 5th IEEE Conference on
  Industrial Electronics and Applications}.\hskip 1em plus 0.5em minus
  0.4em\relax IEEE, 2010, pp. 1675--1680.

\bibitem{bylund2005interactive}
N.~E. Bylund, M.~Andersson, and H.~Knutsson, ``Interactive 3d filter design for
  ultrasound artifact reduction,'' in \emph{IEEE International Conference on
  Image Processing 2005}, vol.~3.\hskip 1em plus 0.5em minus 0.4em\relax IEEE,
  2005, pp. III--728.

\bibitem{ronneberger2015u}
O.~Ronneberger, P.~Fischer, and T.~Brox, ``U-net: Convolutional networks for
  biomedical image segmentation,'' in \emph{International Conference on Medical
  image computing and computer-assisted intervention}.\hskip 1em plus 0.5em
  minus 0.4em\relax Springer, 2015, pp. 234--241.

\bibitem{mathai2019segmentation}
T.~S. Mathai, V.~Gorantla, and J.~Galeotti, ``Segmentation of vessels in ultra
  high frequency ultrasound sequences using contextual memory,'' in
  \emph{International Conference on Medical Image Computing and
  Computer-Assisted Intervention}.\hskip 1em plus 0.5em minus 0.4em\relax
  Springer, 2019, pp. 173--181.

\bibitem{oktay2018attention}
O.~Oktay, J.~Schlemper, L.~L. Folgoc, M.~Lee, M.~Heinrich, K.~Misawa, K.~Mori,
  S.~McDonagh, N.~Y. Hammerla, B.~Kainz, \emph{et~al.}, ``Attention u-net:
  Learning where to look for the pancreas,'' \emph{arXiv preprint
  arXiv:1804.03999}, 2018.

\bibitem{roy2018concurrent}
A.~G. Roy, N.~Navab, and C.~Wachinger, ``Concurrent spatial and channel
  ‘squeeze \& excitation’in fully convolutional networks,'' in
  \emph{International conference on medical image computing and
  computer-assisted intervention}.\hskip 1em plus 0.5em minus 0.4em\relax
  Springer, 2018, pp. 421--429.

\bibitem{hu2018squeeze}
J.~Hu, L.~Shen, and G.~Sun, ``Squeeze-and-excitation networks,'' in
  \emph{Proceedings of the IEEE conference on computer vision and pattern
  recognition}, 2018, pp. 7132--7141.

\bibitem{han2020semi}
L.~Han, Y.~Huang, H.~Dou, S.~Wang, S.~Ahamad, H.~Luo, Q.~Liu, J.~Fan, and
  J.~Zhang, ``Semi-supervised segmentation of lesion from breast ultrasound
  images with attentional generative adversarial network,'' \emph{Computer
  Methods and Programs in Biomedicine}, vol. 189, p. 105275, 2020.

\bibitem{kendall2017uncertainties}
A.~Kendall and Y.~Gal, ``What uncertainties do we need in bayesian deep
  learning for computer vision?'' in \emph{Advances in neural information
  processing systems}, 2017, pp. 5574--5584.

\bibitem{nair2020exploring}
T.~Nair, D.~Precup, D.~L. Arnold, and T.~Arbel, ``Exploring uncertainty
  measures in deep networks for multiple sclerosis lesion detection and
  segmentation,'' \emph{Medical image analysis}, vol.~59, p. 101557, 2020.

\bibitem{kohl2018probabilistic}
S.~Kohl, B.~Romera-Paredes, C.~Meyer, J.~De~Fauw, J.~R. Ledsam, K.~Maier-Hein,
  S.~A. Eslami, D.~J. Rezende, and O.~Ronneberger, ``A probabilistic u-net for
  segmentation of ambiguous images,'' in \emph{Advances in Neural Information
  Processing Systems}, 2018, pp. 6965--6975.

\bibitem{kohl2019hierarchical}
S.~A. Kohl, B.~Romera-Paredes, K.~H. Maier-Hein, D.~J. Rezende, S.~Eslami,
  P.~Kohli, A.~Zisserman, and O.~Ronneberger, ``A hierarchical probabilistic
  u-net for modeling multi-scale ambiguities,'' \emph{arXiv preprint
  arXiv:1905.13077}, 2019.

\bibitem{baumgartner2019phiseg}
C.~F. Baumgartner, K.~C. Tezcan, K.~Chaitanya, A.~M. H{\"o}tker, U.~J.
  Muehlematter, K.~Schawkat, A.~S. Becker, O.~Donati, and E.~Konukoglu,
  ``Phiseg: Capturing uncertainty in medical image segmentation,'' in
  \emph{International Conference on Medical Image Computing and
  Computer-Assisted Intervention}.\hskip 1em plus 0.5em minus 0.4em\relax
  Springer, 2019, pp. 119--127.

\bibitem{meng2019weakly}
Q.~Meng, M.~Sinclair, V.~Zimmer, B.~Hou, M.~Rajchl, N.~Toussaint, O.~Oktay,
  J.~Schlemper, A.~Gomez, J.~Housden, \emph{et~al.}, ``Weakly supervised
  estimation of shadow confidence maps in fetal ultrasound imaging,''
  \emph{IEEE transactions on medical imaging}, vol.~38, no.~12, pp. 2755--2767,
  2019.

\bibitem{amin1989ultrasonic}
V.~R. Amin, ``Ultrasonic attenuation estimation for tissue characterization,''
  1989.

\bibitem{labelme2016}
K.~Wada, ``{labelme: Image Polygonal Annotation with Python},''
  \url{https://github.com/wkentaro/labelme}, 2016.

\bibitem{yongchankwon2018uncertainty}
B.~J.~K. Yongchan~Kwon, Joong-Ho~Won and M.~C. Paik, ``Uncertainty
  quantification using bayesian neural networks in classification: Application
  to ischemic stroke lesion segmentation,'' in \emph{Conference on Medical
  Imaging with Deep Learning}, 2018.

\bibitem{trobaugh1994frameless}
J.~W. Trobaugh, W.~D. Richard, K.~R. Smith, and R.~D. Bucholz, ``Frameless
  stereotactic ultrasonography: method and applications,'' \emph{Computerized
  Medical Imaging and Graphics}, vol.~18, no.~4, pp. 235--246, 1994.

\bibitem{lasso2014plus}
A.~Lasso, T.~Heffter, A.~Rankin, C.~Pinter, T.~Ungi, and G.~Fichtinger, ``Plus:
  open-source toolkit for ultrasound-guided intervention systems,'' \emph{IEEE
  transactions on biomedical engineering}, vol.~61, no.~10, pp. 2527--2537,
  2014.

\bibitem{zu2014orientation}
C.~S. zu~Berge, A.~Kapoor, and N.~Navab, ``Orientation-driven ultrasound
  compounding using uncertainty information,'' in \emph{International
  conference on information processing in computer-assisted
  interventions}.\hskip 1em plus 0.5em minus 0.4em\relax Springer, 2014, pp.
  236--245.

\bibitem{guerrero2007real}
J.~Guerrero, S.~E. Salcudean, J.~A. McEwen, B.~A. Masri, and S.~Nicolaou,
  ``Real-time vessel segmentation and tracking for ultrasound imaging
  applications,'' \emph{IEEE transactions on medical imaging}, vol.~26, no.~8,
  pp. 1079--1090, 2007.

\bibitem{liu2018image}
G.~Liu, F.~A. Reda, K.~J. Shih, T.-C. Wang, A.~Tao, and B.~Catanzaro, ``Image
  inpainting for irregular holes using partial convolutions,'' in
  \emph{Proceedings of the European Conference on Computer Vision (ECCV)},
  2018, pp. 85--100.

\end{thebibliography}

\end{document}